\documentclass[aps,showpacs,prl,twocolumn,superscriptaddress]{revtex4-1}
\usepackage[pdftex]{graphicx}
\usepackage{slashbox}
\usepackage{ulem}

\begin{document}

\title{Strongly momentum-dependent screening dynamics in La$_{0.5}$Sr$_{1.5}$MnO$_4$ observed with resonant inelastic x-ray scattering}
\date{\today}

\author{X. Liu}
\affiliation{Condensed Matter Physics and Materials Science Department, Brookhaven National Laboratory, Upton, New York 11973, USA\\}
\affiliation{Beijing National Laboratory for Condensed Matter Physics, and Institute of Physics, Chinese Academy of Sciences, Beijing 100190, China\\}
\author{T. F. Seman}
\affiliation{Department of Physics, New Jersey Institute of Technology, Newark, New Jersey 07102, USA\\}
\author{K. H. Ahn}
\affiliation{Department of Physics, New Jersey Institute of Technology, Newark, New Jersey 07102, USA\\}
\author{Michel van Veenendaal}
\affiliation{Advanced Photon Source, Argonne National Laboratory, Argonne, Illinois 60439, USA\\}
\affiliation{Department of Physics, Northern Illinois University, De Kalb, Illinois 60115, USA\\}
\author{D. Casa}
\affiliation{Advanced Photon Source, Argonne National Laboratory, Argonne, Illinois 60439, USA\\}
\author{D. Prabhakaran}
\affiliation{Department of Physics, University of Oxford, Clarendon Laboratory, Parks Road, Oxford OX1 3PU, United Kingdom\\}
\author{A. T. Boothroyd}
\affiliation{Department of Physics, University of Oxford, Clarendon Laboratory, Parks Road, Oxford OX1 3PU, United Kingdom\\}
\author{H. Ding}
\affiliation{Beijing National Laboratory for Condensed Matter Physics, and Institute of Physics, Chinese Academy of Sciences, Beijing 100190, China\\}
\author{J. P. Hill}
\affiliation{Condensed Matter Physics and Materials Science Department, Brookhaven National Laboratory, Upton, New York 11973, USA\\}

\begin{abstract}
We report strongly momentum-dependent local charge screening dynamics in CE-type charge, orbital, and spin ordered La$_{0.5}$Sr$_{1.5}$MnO$_4$, based on Mn K-edge resonant inelastic x-ray scattering data. Through a comparison with theoretical calculations, we show that the observed momentum dependence reflects highly localized, nearest neighbor screening of the transient local charge perturbation in this compound with an exciton-like screening cloud, rather than delocalized screening. The size of the screening cloud is estmated to be about 0.4-0.5 interatomic distances.
\end{abstract}


\maketitle

The dynamic screening of the Coulomb interaction plays a central role in determining the electronic properties of materials~\cite{Fetter}. The response of valence electrons to a potential, in particular on time scales of the order of femtoseconds, is through excitation of electron-hole pairs which screen ``bare'' charges in the system. The screening is described theoretically by the density-density correlation function, or its Fourier transform, the dynamic structure factor~\cite{pines}. Spectroscopies that probe the valence band, such as photoemission, are sensitive to these screening dynamics. However, it is often not obvious how to separate the kinetics of a charged particle and the response of the rest of the system to its presence~\cite{photoemission}. In contrast, core-level spectroscopies provide an alternative way of studying the screening dynamics. By removing a deep-lying core electron, a strong local potential is created that exists for a very short time, i.e., the core hole lifetime. Essentially, one creates a short-lived localized ``test'' particle, and measures the response of the electrons to this local transient potential. This type of screening dynamics has drawn the attention of scientists for decades~\cite{Nozieres,vanderLaan,vanVeenendaal}. 

For transition metal compounds, $K$-edge resonant inelastic x-ray scattering (RIXS)~\cite{Ament11} offers the intriguing possibility of projecting the excitations related to the core hole screening onto valence band excitations. Specifically, it has been shown that K-edge RIXS can be directly related to the dynamic structure factor in the limit of a strong or weak core hole potential, $U_{\mathrm{core}}$, relative to the band width~\cite{vandenBrink}. For the case where $U_{\mathrm{core}}$ is comparable to the bandwidth, more typical for $3d$ transition metal compounds, the screening is more complicated because there is an asymmetry between the electron and hole excitations and the intermediate states can not be integrated out~\cite{Ahn09}. In this case, the RIXS response is believed to be sensitive to the transient screening of the intermediate states to the core hole potential~\cite{Ahn09}.

In this letter, we report RIXS measurements of the momentum and energy dependence of the screening dynamics for a transient local potential in a CE-type charge, orbital, and spin ordered manganite, La$_{0.5}$Sr$_{1.5}$MnO$_4$. We find strong momentum dependence of the intensity of the across-gap excitation, with a dramatic increase on moving away from the two dimensional (2D) zone center. We show that this behavior reflects the size and shape of the real-space screening cloud and demonstrate that in La$_{0.5}$Sr$_{1.5}$MnO$_4$, the screening distance is very short, with a screening cloud of about 0.4-0.5 interatomic distances in size.

A single crystal of La$_{0.5}$Sr$_{1.5}$MnO$_4$ was grown by the traveling solvent floating zone method. It has a tetragonal structure at room temperature with {\it I4/mmm} symmetry and undergoes a charge and orbital ordering transition around 230 K, accompanied by complex structural distortions~\cite{Herrero11}. For simplicity, we use here the {\it I4/mmm} notation throughout. The wave vectors of the charge and orbital ordering are then of the form ($\frac{1}{2}$, $\frac{1}{2}$, $L$) and ($\frac{1}{4}$, $\frac{1}{4}$, $L$), respectively. In the low temperature ordered state, La$_{0.5}$Sr$_{1.5}$MnO$_4$ is an insulator with a large gap between the $e_g$ states~\cite{ElecStruc}. The behavior of the excitation between these predominantly Mn $3d$ states, labeled as a $d$-$d$ transition, is the focus of this study. The Mn K-edge RIXS experiments were performed at Advanced Photon Source on beamlines 30-ID and 9-ID with an instrumental energy resolution of about 270 meV (FWHM). The polarization dependence of the RIXS process is controlled by placing the [001] and [110] directions of the crystal in the scattering plane. The incident beam polarization is perpendicular to the scattering plane, i.e., parallel to the [1$\bar{1}$0] direction. Thus the incident polarization condition is fixed for all the ${\bf Q} = (H, H, L)$ points surveyed. All the data presented were collected at $T$ = 20 K, well below the N\'eel temperature(110K)~\cite{TN}. Data are normalized by incident beam intensity and corrected for footprint variations.

\begin{figure}
\includegraphics[width=0.48\textwidth]{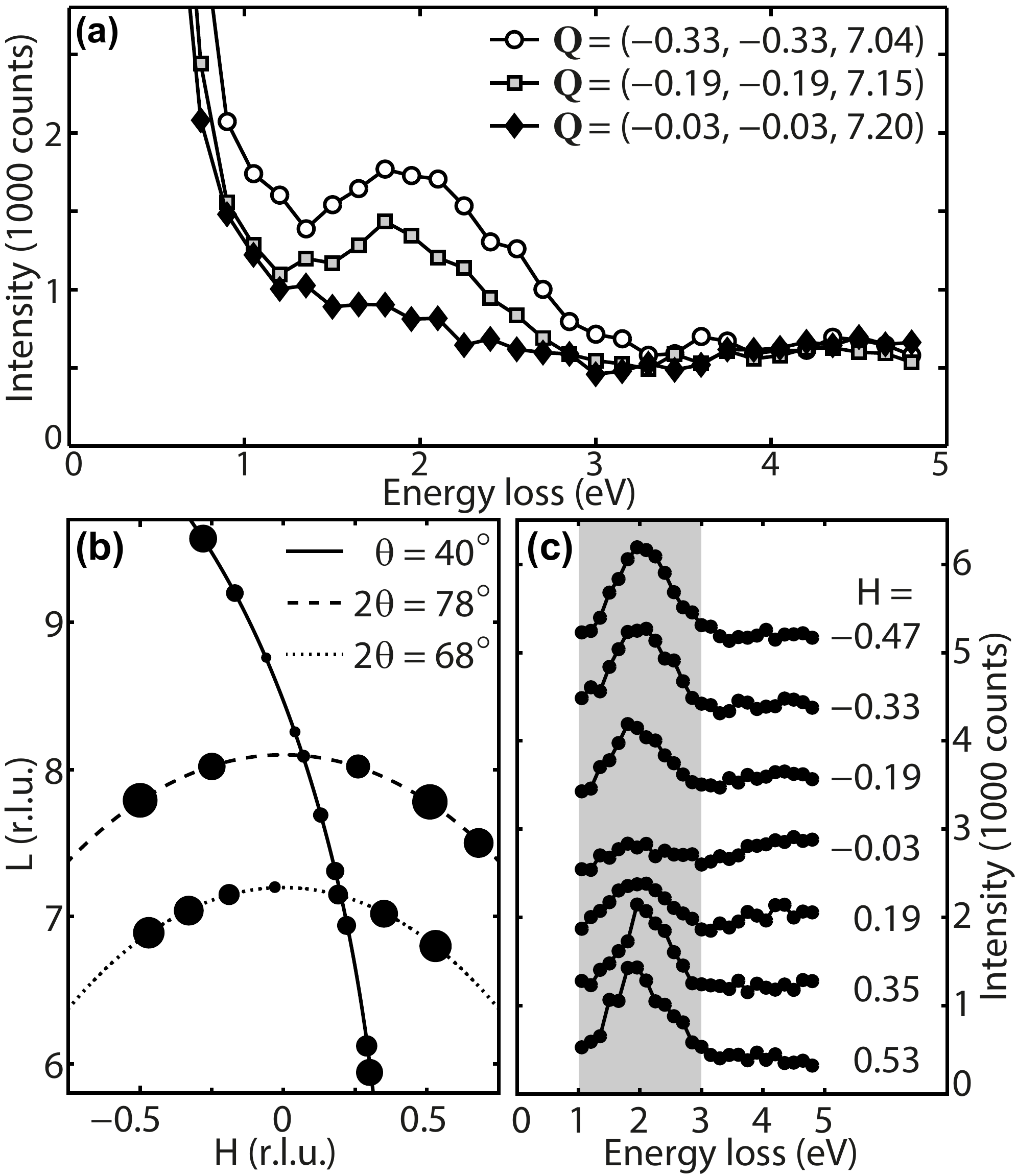}
\caption{(a) RIXS spectra at three {\bf Q} points. (b) The \textbf{Q} points surveyed in the ($H$, $H$, $L$) plane. The radius of the dot is proportional to the integrated intensity of the 2 eV peaks. $\theta$ and $2\theta$ are the incident and detector angles. (c) RIXS spectra for the {\bf Q} points along the $2\theta=68^{\circ}$ line, with the elastic intensity subtracted. The grey shaded region is the energy window used in calculating the integrated intensity of the feature.}
\label{fig:raw-scan}
\end{figure}

In Fig.\,\ref{fig:raw-scan}(a), we show RIXS spectra taken at three {\bf Q} points. In each case, there is a large elastic line centered at zero energy loss. The $d$-$d$ transition appears as a peak on the tail of the elastic scattering at around 2 eV, consistent with optics~\cite{optical} and EELS~\cite{EELS} observations. This across-gap transition has also been observed by K-edge RIXS on other manganites~\cite{RIXS1, RIXS2}. Remarkably, the RIXS spectra show a strong momentum dependence of the intensity of this feature. At ${\bf Q_0} = (-0.03, -0.03, 7.20)$ with very small in-plane momentum transfer, the 2 eV peak is almost unobservable. This momentum dependence is confirmed with RIXS spectra collected at a large number of {\bf Q} points, as shown in Fig.\,\ref{fig:raw-scan}(b) and \ref{fig:raw-scan}(c). To control the systematics resulting from polarization factors, the data were taken either at fixed sample angle, $\theta$, or fixed detector angle, $2\theta$. These conditions result in data taken along three lines in reciprocal space. In all cases, the incident polarization is parallel to the [1$\bar{1}$0] direction. With the detector position $2\theta$ fixed, polarization effects associated with the outgoing x-ray are eliminated. Figure\,\ref{fig:raw-scan}(c) shows RIXS spectra with the elastic line subtracted~\cite{Eline} for the {\bf Q} points with $2\theta$ = $68^{\circ}$. We take the integrated intensity, $I({\bf Q})$, over the $1$-$3$ eV range as a measure of the strength of the 2 eV peak. The size of the symbols in Fig.\,\ref{fig:raw-scan}(b) is proportional to $I({\bf Q})$. A clear systematic dependence on momentum transfer is observed.

\begin{figure}
\includegraphics[width=0.48\textwidth]{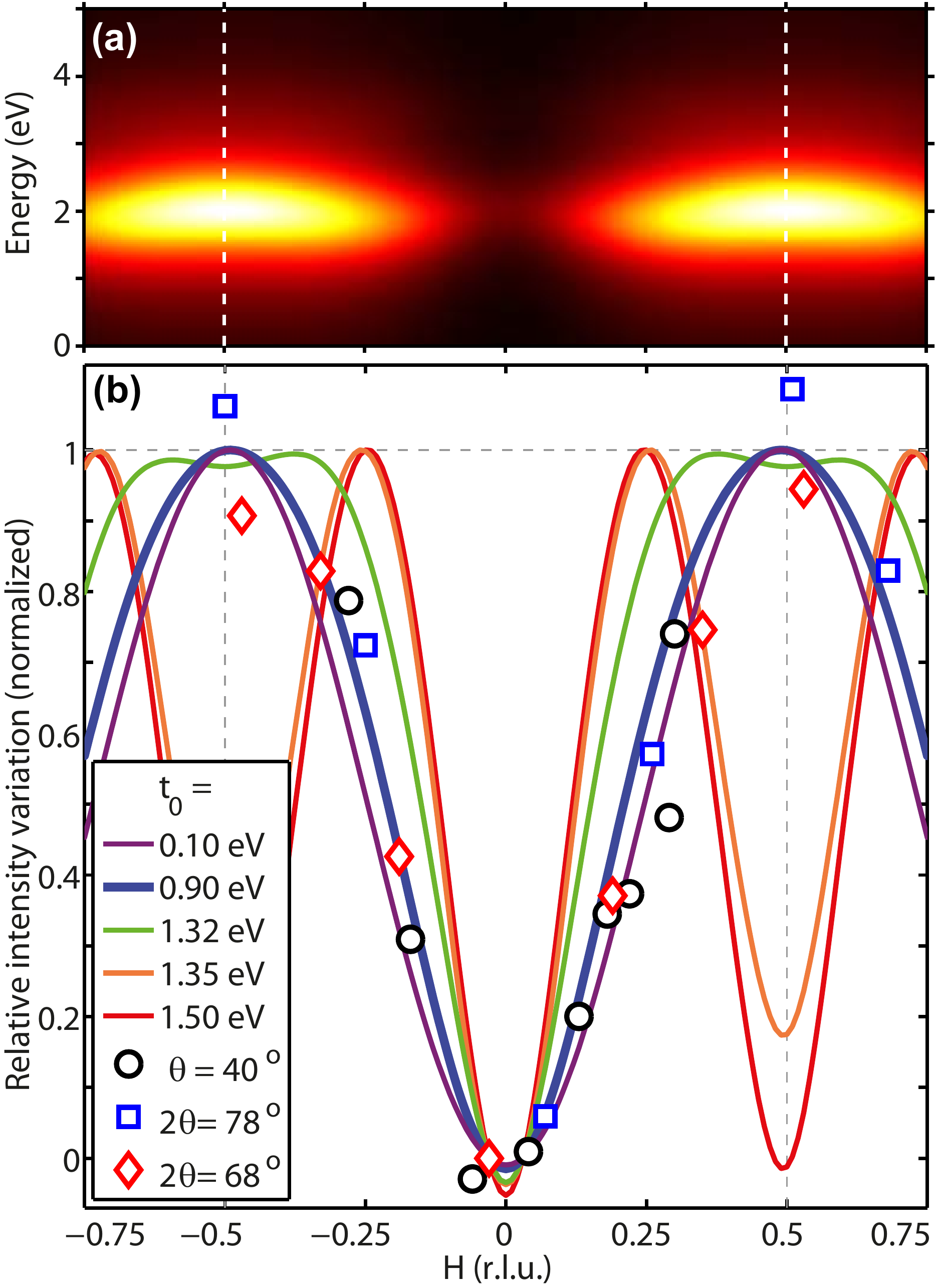}
\caption{(Color online) (a) Contour plot of RIXS intensity calculated for the electron hopping parameter $t_0$ = 0.9 eV, after averaging over twin domains. (b) The integrated RIXS intensity of the 2 eV peak relative to the (-0.03,-0.03) point, plotted with respect to the in-plane momentum transfer along the ($H$, $H$) direction. Symbols represent experimental data. Lines represent theoretical results for different values of $t_0$. Both experimental data and theoretical results are normalized for comparison.}
\label{tCurves}
\end{figure}

The integrated intensities of the $d$-$d$ excitation are plotted as a function of the in-plane momentum transfer in Fig.\,\ref{tCurves}(b). In order to quantitatively compare the experimental data with the theoretical calculations (discussed below), the integrated intensities are plotted relative to the intensity at ${\bf Q_0} = (-0.03, -0.03, 7.20)$~\cite{nearGamma}, i.e., $I({\bf Q})-I({\bf Q_0})$. This removes the uncertainty in determining the common background for all {\bf Q} points. The strength of the 2 eV $d$-$d$ excitation exhibits a minimum at zero in-plane momentum transfer and a maximum at (0.5, 0.5, $L$). Interestingly, although there is a large variation in the $L$ values for the various {\bf Q} points [see Fig.\,\ref{fig:raw-scan}(b)], all the measurements collapse onto a single curve in Fig.\,\ref{tCurves}(b). This demonstrates that there is negligible $L$ dependence to this behavior, a result consistent with the 2D nature of this single layered manganite. Further, it implies that the polarization factors are indeed constant for our experimental geometry. From here on, the momentum transfer will be denoted simply as ${\bf Q}_{2D} = (H, H)$ since the $L$ component is irrelevant.

The experimental data in Fig.\,\ref{fig:raw-scan} and Fig.\,\ref{tCurves} show our main experimental observations. The across-gap $d$-$d$ excitation in La$_{0.5}$Sr$_{1.5}$MnO$_4$, as observed in the RIXS process, exhibits a strong momentum dependence. While the position of the peak shows no appreciable dispersion, the intensity increases rapidly as the in-plane momentum transfer increases away from the 2D zone center. Near the zone center, the spectral weight of the 2 eV feature almost disappears. This is a surprising result. The disappearance of this RIXS spectral weight at the 2D zone center cannot be the result of the dynamic structure factor going to zero, since this feature is still observed in the optical response~\cite{optical}, which probes the zero momentum transfer response function. This demonstrates that K-edge RIXS in La$_{0.5}$Sr$_{1.5}$MnO$_4$ is indeed in the intermediate core-hole potential regime, discussed in the introduction. In the following, we detail this momentum-dependence and show that it arises from the intermediate state screening dynamics and in particular that it reflects the real-space extent of the screening cloud.

To understand this strong in-plane momentum dependence, we calculated the RIXS response from La$_{0.5}$Sr$_{1.5}$MnO$_4$ for a two-dimensional $16 \times 16$ Mn cluster with periodic boundary conditions. The initial and final states of the unperturbed system, and the intermediate states in the presence of the $1s$ core hole on-site Coulomb potential, are solved numerically with a tight-binding approach. The Hamiltonian employed is similar to the one in Ref.~\cite{Ahn00}, which includes the nearest-neighbor electron hopping within the MnO$_2$ plane, the Jahn-Teller and isotropic electron-lattice coupling, the Hund's coupling to the CE-type ordered $t_{2g}$ spins, and the Coulomb interaction between $e_g$ electrons within the Hartree-Fock approximation~\cite{supp}. The sizes of the distortions of the oxygen octahedra are taken from Ref.~\cite{Herrero11}. The RIXS spectra are then calculated from the Kramers-Heisenberg formula~\cite{Ament11,Ahn09}:
\begin{equation}
I \propto \sum_{f}
\left|\sum_{n}\frac{ \langle f| \mathcal{D'}^{\dag}|n \rangle \langle n| \mathcal{D}|g \rangle }
{E_g+\hbar \omega_{\bf k}-E_n + i \Gamma_n}
\right|^2 \delta(E_f-E_g-\hbar\Delta\omega),
\label{eqn1}
\end{equation}
where $|f\rangle$, $|n\rangle$, and $|g\rangle$ represent the final, intermediate and initial states, and $E_f$, $E_n$ and $E_g$ their energies. $\Gamma_n$ is the inverse of the intermediate state lifetime, and $\mathcal{D'}^{\dag}$ and $\mathcal{D}$ are the RIXS dipole transition operators. $\hbar\omega_{\bf k}$ and $\hbar\Delta\omega$ are the incident x-ray energy and the energy loss, respectively. The calculated RIXS intensity is averaged over configurations in which the zig-zag chains of orbital order are along either the [110] or the [1$\bar{1}$0] directions, to take into account twining effects in real crystals. Details of the calculation will be published elsewhere.

The calculated RIXS spectra were found to be most sensitive to the $e_g$-$e_g$ hybridization and the coupling of the $e_g$ electrons to the distortions of oxygen octahedra. These two effects are parametrized as $t_0$ and $\lambda$ in our Hamiltonian, where $t_0$ is the hopping between $3x^2$-$r^2$ orbitals along the $x$ direction and $\lambda$ is proportional to the strength of the electron-phonon coupling~\cite{supp}. With reasonable parameter values~\cite{supp} and the combination of $t_0 = 0.9$ eV and $\lambda = 7.41$ eV/$\AA$, the calculated spectra shown in Fig.\,\ref{tCurves}(a) and the thick (blue) line in Fig.\,\ref{tCurves}(b) closely resemble the experimental observations. The intensity of the calculated RIXS response peaks near 2 eV, and increases rapidly as ${\bf Q}_{2D}$ increases away from (0, 0), towards (0.5, 0.5), as seen in the experiments. We note that the calculated spectra in Fig.\,\ref{tCurves}(a) suggest a slight dispersion of about 130 meV of the 2 eV peak, which is much smaller than that reported for LaSr$_2$Mn$_2$O$_7$~\cite{RIXS2}. Such a small dispersion, roughly equal to the experimental step size taken in Fig.\,\ref{fig:raw-scan}(c), is below the detection limit of our experiment.

The sensitivity of the RIXS response to intersite hopping and the electron-phonon coupling is shown in Fig.\,\ref{tCurves}(b) by varying $t_0$ and $\lambda$~\cite{supp}. For a given $t_0$, $\lambda$ is constrained such that the $d$-$d$ excitation in the RIXS response peaks near 2 eV. Henceforth, only $t_0$ is mentioned for simplicity. The details of the combinations of $t_0$ and $\lambda$ can be found in the supplemental material~\cite{supp}. As was done for the experimental data, the calculated response is integrated over the same 1-3 eV window to generate the curves in Fig.\,\ref{tCurves}(b), and again the value at ${\bf Q_0} = $(-0.03, -0.03) is subtracted. The calculated results show the best agreement with the experimental observations when $t_0 = 0.9$ eV. For larger $t_0$ values, the calculated RIXS response differs significantly from the experimental data. Thus our study sets the upper limit of $t_0$. The inability to precisely determine the parameter values is largely due to the difficulty in determining the contributions from other inelastic scattering processes that give rise to a smooth ``background'' in the low energy loss region. 

\begin{figure}
\includegraphics[width=0.48\textwidth]{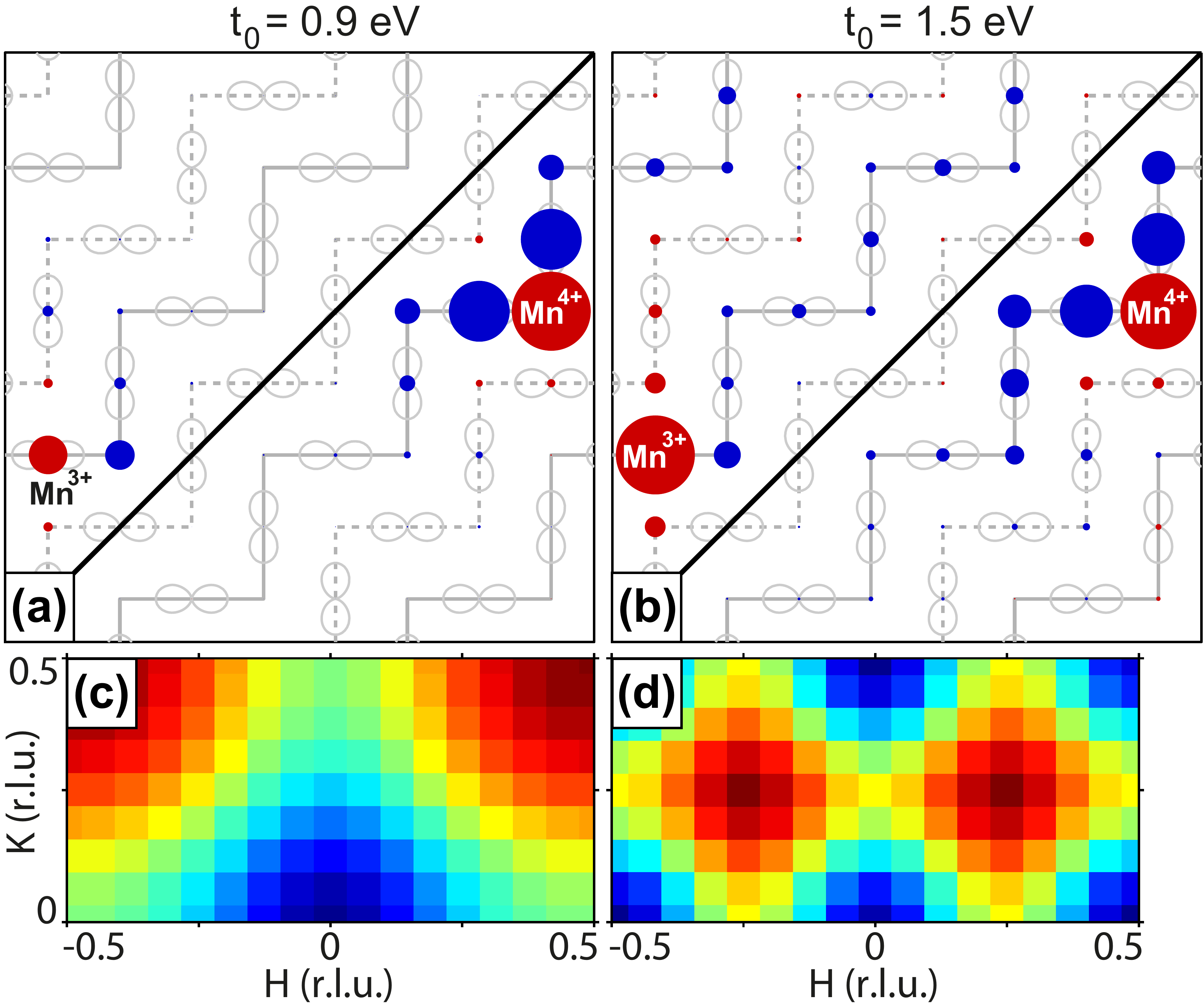}
\caption{(Color online) (a) and (b): Screening configuration in real space for $t_0$ = 0.9 and 1.5 eV, respectively. The top left half corresponds to the case with the core hole at a Mn$^{3+}$ site, while the bottom right it is at a Mn$^{4+}$ site. The volumes of the red and blue spheres are proportional to the electron and hole numbers. The big red spheres at Mn$^{4+}$ core hole sites represent about 0.9 electrons. (c) and (d): Integrated RIXS intensity plotted in the ($H$, $K$, 0) plane of reciprocal space for $t_0$ = 0.9 eV and $t_0$ = 1.5 eV, respectively. Red and blue represent the maximum and minimum intensities, respectively.}
\label{fig:rMap}
\end{figure}

To further understand the implications of the observed momentum dependence of the RIXS spectrum for the screening dynamics, we calculate the real-space screening configurations from the lowest energy intermediate eigenstates. These are shown in the top panels of Fig.\,\ref{fig:rMap}. Figures \ref{fig:rMap}(a) and \ref{fig:rMap}(b) compare the charge redistributions for $t_0$ = 0.9 and 1.5 eV, respectively, with the core hole at either a Mn$^{3+}$ or a Mn$^{4+}$ site. The volumes of the red and blue spheres scale with the screening electron and hole densities on the individual sites. Figures \ref{fig:rMap}(c) and \ref{fig:rMap}(d) show the calculated RIXS intensities for the two $t_0$ values over half of a Brillouin zone.

As expected, the excited hole distributions are more localized near the core hole sites for the smaller value of $t_0$. For $t_0 = 0.9$ eV, the screening hole is tightly bound to the excited electron with more than 90 \% of the excited charge located on the three nearest neighbor sites along the zig-zag chain. The predominant wave vector for these electron and hole distributions is (0.5, 0.5), coincident with the location of the RIXS peak intensity maximum in reciprocal space in Fig.\,\ref{fig:rMap}(c) and in agreement with our experimental results. For the large hopping parameter, $t_0 = 1.5$ eV, the screening pattern in real space changes drastically. The majority of the hole distribution in Fig.\,\ref{fig:rMap}(b) is now beyond the nearest neighbor sites, and is spread throughout the zig-zag chains. This difference in screening dynamics is directly reflected in the RIXS response, with the maximum of the RIXS response then shifted to  around (0.25, 0.25), as shown in Fig.~\,\ref{tCurves}(b) and Fig.\,\ref{fig:rMap}(d). This pattern is completely at odds with that seen in the experiment results.

\begin{figure}
\includegraphics[width=0.48\textwidth]{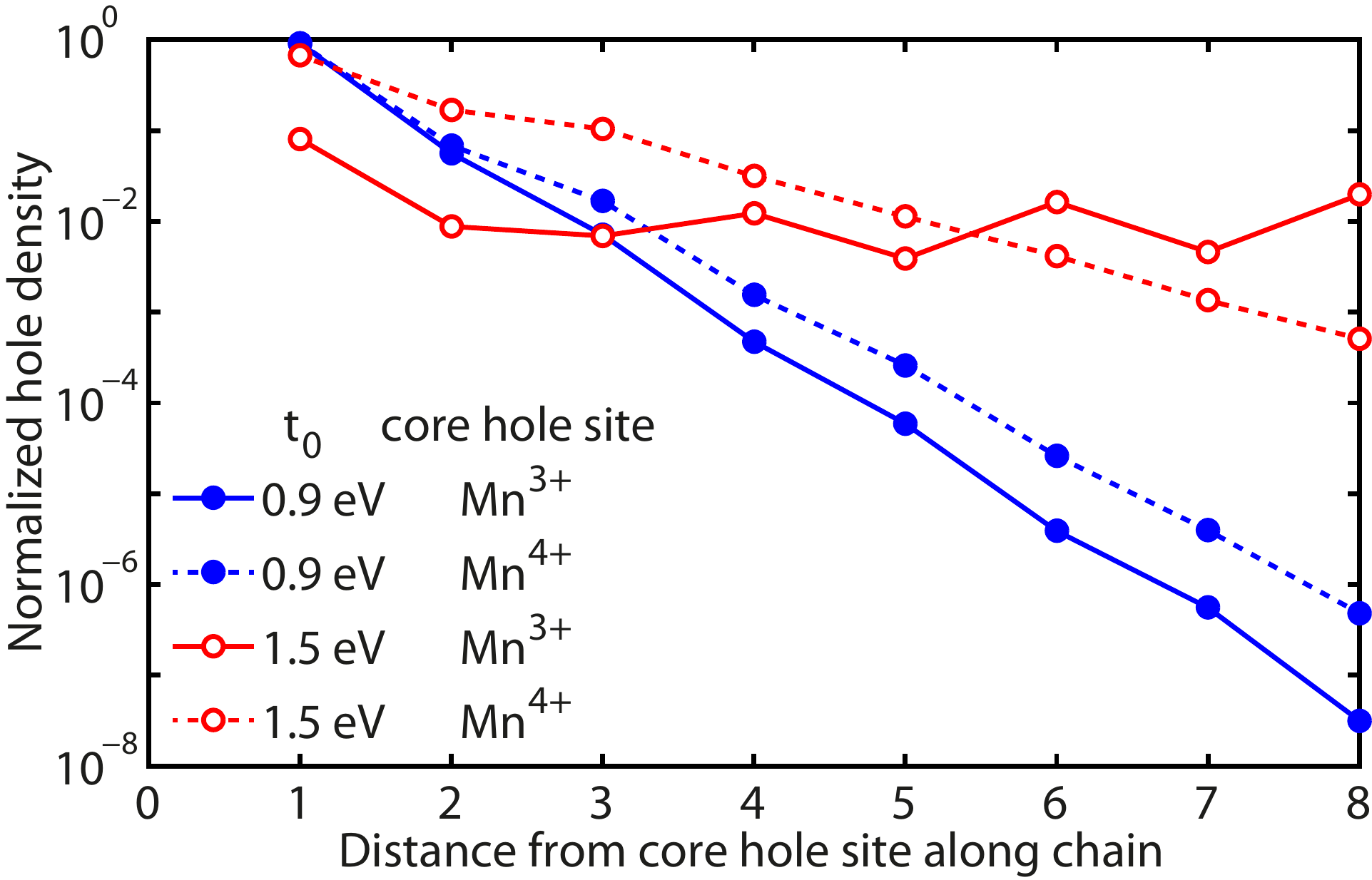}
\caption{(Color online) The excited hole number normalized by the excited electron number at the core hole site, plotted in semi-logarithmic scale with respect to the distance from the core hole site along the zig-zag chain.}
\label{fig:Chain}
\end{figure}

The relationship between the hopping strength and the charge redistribution in the screening process apparent in the theoretical calculations is shown more clearly in Fig.\,\ref{fig:Chain}, where the relative hole number at a given site is plotted as a function of the distance from the core hole site, as measured along the zig-zag chain. The semi-logarithmic plot reveals an exponential decay of the hole density, confirming the presence of exciton-like screening clouds. The size of the screening cloud, which characterizes the screening dynamics and determines the RIXS response, depends strongly on the hopping strength $t_0$. Taking the $t_0 = 0.9$ eV case, which best describes the RIXS data, we fit the excited hole distributions to an exponential function and find that the size of the screening clouds are 0.4 and 0.5  atomic spacings for the Mn$^{3+}$ and Mn$^{4+}$ sites, respectively.

In summary, we observe a highly momentum-dependent K-edge resonant inelastic x-ray scattering intensity
in the orbital ordered, layered manganite La$_{0.5}$Sr$_{1.5}$MnO$_4$. We interpret this through a comparison with calculations based on a tight-binding approach, and show that these observations imply a highly localized, nearest neighbor screening of the local charge perturbation. We further find that the momentum dependence of the RIXS spectrum reflects the pattern and range of the screening in real space, and thus are able to measure the size and shape of the screening cloud. We find that the screening cloud is localized to a few Mn sites in the Mn-O plane, emphasizing the short range nature of the Coulomb interactions in the manganites. 

The work at Brookhaven was supported by the U.S. Department of Energy, Division of Materials Science, under Contract No. DE-AC02-98CH10886. X.L. is also supported by the Institute of Physics, Chinese Academy of Science. M.v.V. was supported by the U. S. Department of Energy, Office of Basic Energy Sciences, Division of Materials Sciences and Engineering under Award No. DE-FG02-03ER46097 and NIU Institute for Nanoscience, Engineering, and Technology. The collaboration between T.F.S., K.H.A., and M.v.V. were supported by the Computational Materials and Chemical Science Network under Grant DE-FG02-08ER46540 and Grant DE-SC0007091. K.H.A. is further supported by Argonne X-ray Science Division Visitor Program. Work at Argonne National Laboratory and Use of the Advanced Photon Source was supported by the U. S. DOE, Office of Science, Office of Basic Energy Sciences, under contract No. DE-AC02-06CH11357. H.D. was supported by Grant 2010CB923000 from MOST, China. Work in Oxford was supported by the U.K. Engineering and Physical Sciences Research Council.



\begin{thebibliography}{99}
\bibitem{Eline} The elastic line was subtracted by using a model function. The latter was determined by measuring the elastic scattering at an off-resonant energy (7 eV below the edge), where there are negligible inelastic contributions.
\end{thebibliography}
\end{document}